\documentclass[12pt]{article}  
\usepackage{graphicx}  
\usepackage{dcolumn}   
\usepackage{bm}        
\usepackage{amssymb}   
\usepackage{amsmath}   
\usepackage{cite}
\usepackage{color}
\usepackage{float}
\def\beq{\begin{equation}}
\def\eeq{\end{equation}}

\def\bea{\begin{eqnarray}}
\def\eea{\end{eqnarray}}
\def\ba{\begin{array}}                  
\def\ea{\end{array}}

\newcommand{\brat}{{\rm BR}}
     
\begin{document}

\begin{center}
\noindent{\Large \bf Anapole moment of the lightest neutralino 
in the cMSSM}  
\\[15pt]
{\large Luis G. Cabral-Rosetti$^a$,  Myriam
  Mondrag{\'o}n$^b$,\\ Esteban Reyes-P\'erez$^b$}\\[15pt]
{\small 
 $^a$Centro Interdisciplinario de Investigaci\'on y Docencia en 
Educaci\'on T\'ecnica, CIIDET. Av. Universidad 282 Pte., Col. Centro,
A. Postal 752, C.P. 76000, Santiago de Quer\'etaro, Qro., M\'exico.
\\[3pt]
  \texttt{cabralrosetti@gmail.com}  \\[7pt]
$^b$  Instituto de F{\'\i}sica, Universidad Nacional Aut\'onoma de M{\'e}xico\\
  Apdo. Postal 20-364, C.P. 01000, Ciudad de M{\'e}xico, M{\'e}xico.\\[3pt]
  \texttt{myriam@fisica.unam.mx}\quad  
\texttt{esteban.reyesperez@gmail.com}
}
\end{center}

\begin{abstract}
  We study the anapole moment of the lightest neutralino in the
  constrained Minimal Supersymmetric Standard Model (cMSSM). The
  electromagnetic anapole is the only allowed electromagnetic form
  factor for Majorana fermions, such as the neutralino. Since the
  neutralino is the LSP in many versions of the MSSM and therefore a
  candidate for dark matter, its characterization through its
  electromagnetic properties is important both for particle physics
  and for cosmology. We perform a scan in the parameter space of the
    cMSSM and find that the anapole moment is different from zero
    albeit very small ($< 10^{-3}$ GeV$^{-2}$). Combined with
    experimental constraints like the Higgs mass and the DM relic
    density, the allowed region of parameter space lies within the
    reach of future direct DM searches. Thus, the anapole moment could be used as a complementary constraint when studying the parameter space of the cMSSM and other similar models.

\end{abstract}


\section{Introduction}

One of the best motivated extensions of the Standard Model (SM) is the Minimal
Supersymmetric Standard Model (MSSM), since, besides giving a solution
to the hierarchy problem, it provides us with a good candidate for cold
dark matter (CDM), namely, the lightest neutralino.

A number of different experiments are working (or will be soon) in the
search for a direct or an indirect signal of dark matter (DM) (for
recent reviews on dark matter detection see
\cite{Szelc:2010zz,Schnee:2011ef}).  If DM is detected, there will be
a need to differentiate between the candidates, characterizing them as
much as possible. In the last few years, there has been intense work
on the electroweak properties of dark matter since they might be
relevant for the calculation of DM decays and annihilations,
\cite{Sigurdson:2004zp,Ciafaloni:2010ti,Ciafaloni:2010qr,Ciafaloni:2011sa,Bell:2010ei,
  Bell:2011eu,Bell:2011if,Dent:2008qy,Kachelriess:2009zy,Heo:2009xt,Heo:2009vt,Liebler:2010bi,Barger:2010gv},
which have consequences in astrophysical processes
\cite{Berezinsky:2002hq,Kachelriess:2007aj} and are important in
indirect astrophysical searches for DM, as in the calculation of the
annihilation cross section of the DM itself.

Motivated by this, since 2009 we have been studying the
  toroidal dipole moment (TDM) of Majorana particles, which is related
  to the anapole moment, one of the least studied electromagnetic
  properties of a particle
  \cite{CabralRosetti:2009zz,CabralRosetti:2011zz,CabralRosetti:2012tb,Cabral-Rosetti:2014cpa}.
  Lately, there has been a surge of interest in the study of anapole
  moments from the astrophysical as well as the particle physics points 
  of view (e.g. \cite{Ho:2012bg,Kopp:2014tsa,DelNobile:2014eta,Ibarra:2016dlb}).

Since first discussed in 1957, the anapole moment, introduced by
Zel'dovich \cite{Zeldovich:1957}, has been  investigated in
different fields of science and technology. The anapole moment corresponds to a $T$ invariant interaction, which
is $C$ and $P$ non-invariant \cite{Zeldovich:1957}. Within the Standard Model
it has been calculated in neutrino and hadron physics
\cite{Dubovik:1975ch,Abak:1987nh,Dubovik:1996gx,Bukina:1998wn}.  In
nuclear physics it has been studied in atomic nuclei
\cite{Sahoo:2015qcd,Stadnik:2013raa,Roberts:2014cga}.  In relation to
the DM problem, there are different proposals to observe this
``physical observable'', extracting its value from direct measurements
between DM and atomic nuclei
\cite{Stadnik:2014xja,Kopp:2014tsa,DelNobile:2014eta,Gresham:2014vja}.
Finally, in engineering various aplications have been studied, in
areas such as ``Ferrite resonators'' \cite{Kamenetskii:2011ds},
electromagnetic properties in dielectric nanoparticles
\cite{miroshnichenko2015nonradiating}, and in the study of
electromagnetic radiation in antennae with a helicoidal toroidal
geometric distribution
\cite{1971AmJPh..39.1039P,Corum:1987,1990JPhA...23.5755A,Afanasev:1991mb,1995AmJPh..63..717C},
among others.

Recently, Ho and Scherrer have proposed that dark matter interacts
with ordinary matter exclusively through the anapole moment
\cite{Ho:2012bg}.  They calculate the anapole moment needed to obtain
the right amount of DM relic abundance, and the anapole DM signatures
that could be observed in the LHC \cite{Gao:2013vfa}. 
Haish and Kahlhoefer have shown the importance of
loop contributions to the scattering cross section of dark matter, in
particular those induced by the anapole interaction
\cite{Haisch:2013uaa}. More recently, del Nobile et al made a
halo-independent analysis of direct DM detection data considering that
it has only anapole and magnetic moment dipole interactions
\cite{DelNobile:2014eta}. 
Also, an analysis on the loop corrections for leptophilic DM and
internal bremsstrahlung was presented in \cite{Kopp:2014tsa}, where
the authors calculate the DM anapole and dipole moments in a toy model
using direct detection data.

In this paper we calculate the anapole moment of the neutralino at the
one-loop level within the constrained Minimal Supersymmetric Standard
Model (cMSSM).  In the MSSM the anapole contributions of the neutralino arise
exclusively through radiative corrections to the vertex $\chi
\bar{\chi} \gamma$. We do a scan in the five parameter space of the
  cMSSM and compare the results for the anapole moment with the
  above mentioned experimental limit.  In our analysis we take into
  account also other experimental constraints, namely the Higgs boson
  mass and the decays $b\to s\gamma$ and $B\to \mu^+ \mu^-$ in order to
  find the viable regions of parameter space.  We find that although
  the anapole moment is very small througout the regions studied, it is possible to distinguish
  between the different regions of parameter space through it, which
  makes it indeed  an important property  when characterising dark
  matter. Our results agree qualitatively with those found by Ho and
  Scherrer \cite{Ho:2012bg}. 

The article is organised as follows: in section II we present a very
brief summary of some aspects of the constrained MSSM relevant
to our calculation.  In section III we review the general form for the
electromagnetic vertex of a particle, and in particular for a Majorana
particle. We introduce the anapole moment and its relation to the
toroidal dipole moment. In section IV we explain the methodology used
to calculate the anapole moment of the neutralino in the cMSSM and
evaluate it for different values of the parameters. Section V presents
the obtained results and our conclusions.

\section{The MSSM and the neutralino as candidate for dark matter}

According to the latest results of WMAP, the CDM density is \cite{Larson:2010gs}
\beq
\Omega_{DM}h^2 \sim 0.1109 ~,
\eeq
where $h$ is the Hubble constant in units of $100$ km sec$^{-1}$ Mpc$^{-1}$.
The thermally averaged effective cross section times the relative
speed of the dark matter particle, needed to get this relic density is
\cite{Jungman:1995df,Olive:2003iq,Feng:2010gw}
\beq
<\sigma v> \propto g^4_{weak} /16\pi^2 m_x^2
\eeq
consistent with the assumption of a weakly interacting dark matter
particle (WIMP) with a mass between $10 ~$GeV - $(few)$ TeV.

The minimal supersymmetric extension of the Standard Model (MSSM)
provides us with one of the best WIMP candidates for dark matter: the
lighest neutralino (for reviews on SUSY see for instance
\cite{Martin:1997ns, Aitchison:2005cf}).  The MSSM
requires two complex Higgs electroweak doublets to give mass to the up and
down type quarks in order to avoid chiral anomalies. 
The MSSM has also a new discrete symmetry, R parity, defined as 
 $R = \left( -1\right)^{3B+2S+L}$, where $B$ and $L$ are the baryonic
 and leptonic numbers respectively. This symmetry assignes a charge
 +1 to the SM particles and -1 to the supersymmetric partners,
 making the lightest supersymmetric particle (LSP) stable. 

 However, supersymmetry has to be broken or it would have already been
 observed. To break it explicitly without the reappearance
 of quadratic divergencies, a set of super-renormalizable terms are
 added to the Lagrangian, the so-called soft breaking terms. The
 Lagrangian for the soft breaking terms is given by
\begin{equation} {\cal L}_{soft} = -\frac{1}{2}M_a\lambda^a\lambda^a
  -\frac{1}{6}A^{ijk}\phi_i\Phi_j\phi_k -
  \frac{1}{2}B^{ij}\phi_i\phi_j + 
  c.c. -(m^2)^i_j\phi^{j*}\phi_i ~,
\end{equation}
 where $M_a$ are the gaugino masses, $A^{ijk}$ and $B^{ij}$ are
trilinear and bilinear couplings respectively, and $(m^2)^i_j$ are
scalar squared-mass terms.  It is assumed that supersymmetry breaking
happens in a hidden sector, which communicates to the observable one
only through gravitational interactions, and that the gauge
interactions unifiy. This means that at the GUT scale the soft
breaking terms are ``universal'', i.e., 
the gauginos
$M_a$ have a common mass, as well as the scalars $(m^2)^i_j$ and the trilinear
couplings, $A^{ijk}$. Requiring electroweak symmetry breaking fixes
the value of $B^{ij}$ and the absolute value of the Higgsino mixing
parameter $|\mu|$. This is known as the constrained MSSM (cMSSM) which
is described by five parameters: the unified gaugino mass
$m_{1/2}$, the universal scalar mass $m_0$, the value of the universal
trilinear coupling $A_0$, the sign of Higgsino mass parameter $\mu$, and the
ratio of the vacuum expectation values of the two Higgses, $\tan
\beta$.

After the electroweak symmetry breaking the neutral and charged states
in the MSSM can mix.  In the case of the neutral ones they give rise to
a set of four mass eigenstates, the neutralinos.  It is the lightest
one of these that is the LSP and a good candidate to dark matter in
many SUSY models.  The lightest neutralino, in the gauge eigenstate
basis, is a function of the neutral higgsinos and the neutral
gauginos (wino and bino) and its properties will depend on the mixing, which in turn depends on the soft breaking parameters.

Before WMAP, the cMSSM was compatible with the limit $\Omega _{DM,0}
h^2\sim 0.1 - 0.3$ and other direct and indirect low energy and
collider data in a huge parameter space region called the
``bulk''. However, after the constraint by WMAP to $\Omega _{DM,0}
h^2$,  and with the recent limits to the
sparticles masses from the LHC, which excludes light masses, the
bulk region in $m_0 - m_{1/2}$ is no longer viable. Moreover, in the
cMSSM the LSP neutralino is, practically for all cases, an almost pure
bino state which annihilates itself more efficiently into leptons
through right hand sleptons due to their higher hypercharge. However,
with the newest data from WMAP, this mechanism is not sufficiently
efficient. There are still three favoured scenarios that require some
very specific accidental relations between some parameters at the
electroweak scale.

In the cMSSM at low $m_0$, there is a region with almost degenerate
$\tilde{\tau} - \tilde{\chi}^0_1$. In this case the populations of
these two particles are almost the same, making the NLSP
$\tilde{\tau}$ thermally accesible.  The mass diference
  between the scalar tau and the lighest neutralino, $\Delta M =
  m_{\tilde{\tau}}-m_{\tilde{\chi}^0_1}$, 
  controls the population ratio of these two species through the
  Boltzmann factor $\exp (-\Delta M/T_f)$. Therefore, it is a very
  sensitive parameter which enters into the calculation of the relic
  density. Whenever the coannihilation takes place, through the
  participation of $\tilde{\tau}$ in processes like $\tilde{\tau}_1
  \tilde{\chi}^0_1 \longrightarrow \tau \gamma$ or even
  $\tilde{\tau}_1\tilde{\tau}_1 \longrightarrow \tau \tilde{\tau}$,
  the relic density can be reduced in comparison to the case of the
  bulk scenario. In this region, the LSP neutralino is mainly bino
  with a mass esentially stablished by $M_1$ up to corrections of
  order $M^2_Z/\mu$ ($\mu$ is high). The approximate formulae for the
  mass of the neutralino and the mass of $\tilde{\tau}_1$ suggest that
  degeneration happens for $m_0 \sim 0.145 ~ m_{1/2}$.

A sudden increase in the usual mechanism of coannihilation to reduce
the relic density can occur if $m_{\tilde{\chi}^0_1}$ is close to a
pole. Faster and more efficient annihilation can take place through
Higgs resonance. Given the Majorana nature of the neutralino, the
resonant \textit{enhancement} is obtained only via the pseudo-scalar
Higgs boson. The colliders constrains to the LSP in the cMSSM allow
the heavy ``Higgs funnel", where $\tilde{\chi}^0_1\tilde{\chi}^0_1
\longrightarrow A \longrightarrow b\bar{b}/\tau \bar{\tau}$, which
happens for high $\tan \beta$. Therefore, the quantities that
establish this scenario are the quantity $2m_{\tilde{\chi}^0_1}-m_A$
and the amplitude of the pseudo-scalar, since they define the
resonance profile of $A$.

In most of the cMSSM, $\mu$ is very high. However, you can
exceptionally have that $\mu \sim M_1$, which allows much more
efficient coannihilation through reactions as
$\tilde{\chi}^0_1\tilde{\chi}^0_1 \longrightarrow
WW/ZZ/Zh/t\bar{t}$. This happens in the so-called focus point region
where $m_0$ is very high. The focus point region corresponds to high
values of $m_0$ close to the border of viable electroweak symmetry
breaking, where the value of $\mu$ decreases rapidly. When $\mu \sim
M_1$, $M_2$, the LSP has a significant fraction of higgsino, and the
next lightest sparticles ($\tilde{\chi}^0_2$ o $\tilde{\chi}^{\pm}_1$)
have also a significant component of higgsino and are not much heavier
than the LSP. Thus, the coannihilation channels are favoured. However,
coannihilation cannot be very efficient, otherwise the relic density
would be less than what is actually measured. In this scenario, all
the sfermions are very heavy (more than $4$ TeV) to be accesible to
one of the proposed colliders. The LSP mass goes from close to 150 to
350 GeV, with higgsino-type neutralinos being 100 to 50 GeV
heavier. From the perspective of a linear collider,  an energy
of more than 800 GeV is needed to reveal some of the properties of this
scenario. The pseudo-scalar has a mass higher than 1 TeV and very
likely would not be found directly in the LHC.

There have been several scans of the cMSSM and other MSSM type models
using different criteria to check its viability.  Many of these are
either Bayesian
(e.g. \cite{Roszkowski:2014wqa,Fowlie:2014faa,Balazs:2013qva,
Strege:2012bt,Cabrera:2012vu}) or frequentist
(e.g. \cite{Buchmueller:2012hv,Bechtle:2014yna,Bagnaschi:2015eha}),
or both (\cite{Bechtle:2012zk,Strege:2011pk}), but there are also many
studies using other techniques or constraints to perform the analysis
(e.g. \cite{Abdallah:2015hza,Catalan:2015cna,Ellis:2015rya,Allanach:2013cda,Allanach:2013yua}). In
the cMSSM it is assumed that DM is composed by only one type of
particle, the LSP, which is usually assumed to be the neutralino. Most
of the studies then look for regions of likelihood with different
boundary conditions at the GUT scale, i.e. values for $m_0, m_{1/2},
\tan\beta, \mu$ and $A_0$, and phenomenological constraints at low
energies, which usually include a combination of the following: the
branching ratios $b\to s\gamma$, $B\to \mu^+\mu^-$, the dark matter
relic density, the requirement of radiative electroweak symmetry
breaking, the Higgs mass, and a solution to the $g-2$ problem
(although not all of these constraints may be considered together in
every analysis), and constraints coming from direct and indirect
searches for dark matter.  From these various scans it is clear that
the cMSSM is highly challenged, although there are still regions of
parameter space allowed, depending on which low energy constraints are
used.  In general, to satisfy most of the above constraints, a heavy
supersymmetric spectrum is expected, with $A_0 \neq 0$ and large
$\tan\beta$ (for a recent overview on constraints on the cMSSM and
other SUSY models see for instance
\cite{Buchmueller:2012hv,Buchmueller:2013psa,Bagnaschi:2015eha,Ellis:2015cva}).

\section{Anapole Moment}

For $1/2$-spin particles the most general expression for the
electromagnetic vertex function, which characterizes the interaction
between the particle and the electromagnetic field, is:
\begin{eqnarray}
\Gamma _\mu  (q) = f_Q (q^2 )\gamma _\mu   + f_\mu  (q^2 )i\sigma _{\mu \nu } q^\nu\gamma _5 \nonumber \\   - f_E (q^2 )\sigma _{\mu \nu } q^\nu   + f_A (q^2 )(q^2 \gamma _\mu   - \displaystyle{\not}q q_\mu )\gamma _5 ,
\end{eqnarray}
where $f_Q (q^2)$, $f_\mu (q^2)$, $f _E (q^2)$ and $f _A(q^2)$ are the
so called charge, magnetic dipole, electric dipole and anapole form
factors, respectively; in here $q _{\mu} = p _{\mu} ' - p_{\mu}$ is
the transferred 4-momentum and $\sigma _{\mu \nu} = (i/2) \left[
  {\gamma _\mu ,\gamma _\nu } \right]$ \cite{Dubovik:1996gx,
  Bukina:1998kw}. These form factors are physical observables when
$q^{2}\rightarrow0$, and their combinations define the well known
electric charge (Q), magnetic dipole $(\mu)$, electric dipole $(d)$
and anapole $(a)$ moments.

However, the electromagnetic properties of the neutralino (which is a Majorana particle) are described by a unique form factor, the anapole, $f_A
(q^2)$. This is a consequence of CPT-invariance and the C, P, T
properties of $\Gamma _\mu (q^{2})$ and the interaction
Hamiltonian. Thus, the electromagnetic vertex function of the neutralino
can be written just as
\begin{eqnarray}
  \Gamma _\mu  (q^2) =  f_A (q^2 )(q^2 \gamma _\mu   
- \displaystyle{\not}q q_\mu )\gamma _5 .
\label{eq:electro-neutralino}
\end{eqnarray}  

The anapole moment was introduced by Zel'dovich to describe a
T-invariant interaction that does not conserve P and C parity
\cite{Zeldovich:1957}. In contrast to the electric and magnetic dipole
moments, the anapole moment interacts only with external
electromagnetic currents $J_{\mu} = \partial^{\nu} F_{\mu \nu}$.
 In the non-relativistic limit, the
interaction energy with an external electromagnetic field takes the
form
\begin{equation}
{\cal H} _{int} \propto -\mu \left( \sigma \cdot B \right) - d\left(\sigma \cdot E\right)  -a \left( \sigma \cdot \bigtriangledown \times B\right),
\end{equation}
where $B$ and $E$ are the strength of the magnetic and electric fields, and $\vec{\sigma}$ are the Pauli spin matrices.

The anapole moment does not have a simple classical analogue, since
$f_A(q^2)$ does not correspond to a multipolar distribution. A more
convenient quantity to describe this interaction was proposed by
V. M. Dubovik and A. A. Cheshkov \cite{Dubovik:1975ch}: the toroidal
dipole moment (TDM), $T(q^2)$. For a comprenhensive review on complete electromagnetic
multipole expansions, including toroidal ones, see \cite{Gongora:2006lk}.  The TDM and the anapole moment
coincide in the case of $m_i = m_f$, i.e. the incoming and outgoing
particle are the same. This type of static multipole moments does not
produce any external fields in vacuum but generate a free-field (gauge
invariant) potential \cite{Dubovik:1996gx}, which is responsible for
topological effects like the Aharonov-Bohm one.

\begin{figure}
  \centering
  \includegraphics[height=.25\textheight]{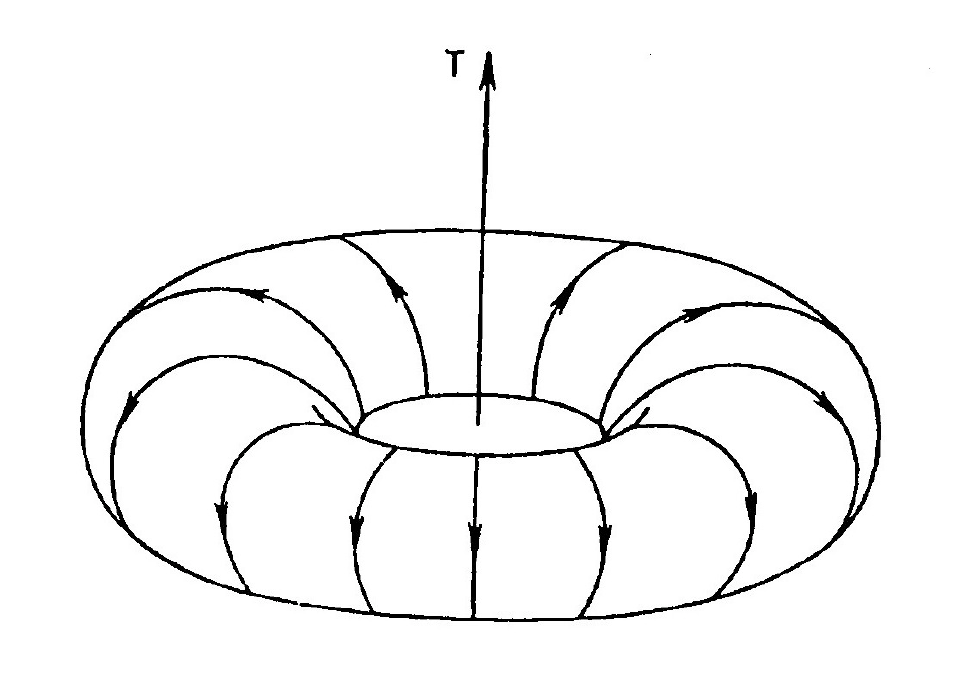}
  \caption{Current configuration with a toroidal dipole moment. The
    arrows on the torus indicate the direction of the current, and the
    TDM is directed towards the axis of symmetry of the torus.}
\label{toro}
\end{figure}

The simplest TDM model (anapole) was given by Zel'dovich as a
conventional solenoid rolled up in a torus and with only one poloidal
current, see Fig. \ref{toro}. For such stationary solenoid, without
azimuthal components for the current or the electric field, there
is only one magnetic azimuthal field different from zero inside
the torus.

As mentioned in the introduction, the anapole moment is a very useful
quantity in nuclear physics, where it has been widely studied, as well
as in astrophysics and engineering.  In particle physics it is important in DM
detection, since the DM candidates can have couplings to nuclear
spins. There are also limits to
detection of anapole dark matter in the LHC \cite{Gao:2013vfa}, which
exclude it for masses $\lesssim 100$ GeV.

To measure the anapole moment of DM, direct detection is needed, where
the resulting cross section from the scattering of the DM particle
with a nucleus is measured, and from there the anapole moment or bounds
to its value can be extracted. The first such upper bound was calculated in
ref.~\cite{Pospelov:2000bq}, using data from the DAMA/LIBRA collaboration
\cite{Bernabei:1996vj} and from the Ge detector of the CDMS 
collaboration \cite{Abusaidi:2000wg}
\begin{equation}
\sim 4 \times 10^{-2} \rm{fm}
\end{equation}
for a WIMP mass of 100 GeV (3 for the Ge detector and 4 for the NaI one).

There are currently several experiments exploring direct DM detection
(for a recent review on indirect and direct DM searches see
\cite{Klasen:2015uma}).  The best exclusion bounds for DM at present
come from XENON100 \cite{Aprile:2012nq} and LUX \cite{Akerib:2013tjd}.
Although the anapole moment is a very small quantity, it is expected
that the improvement in the sensitivity of future direct DM detection
experiments will allow to put more stringent bounds on its value.  In
this respect, knowing precisely the neutron and proton spin contents
of relevant nuclei is important for the correct interpretation of the
data (see for instance
\cite{Stadnik:2014xja,Roberts:2014cga,Gresham:2014vja}, and references
therein).  It is expected that both XENON1T \cite{Davide:2015bca} and
LUX-ZEPLIN  (LZ)\cite{Akerib:2015cja} will improve by $\sim 100$ times
their measurement of the WIMP-nucleon cross section. In the case of
XENON1T the sensitivity is expected to be $2\times 10^{-47}$~cm$^2$ at
a WIMP mass of $40\sim 50$ GeV \cite{Kessler:2014jya,Davide:2015bca},
whereas LZ has a projected sensitivity of
$10^{-48}$~cm$^2$ for its full 1000 day exposure
\cite{Dobson:2014rxa,Akerib:2015cja}.

\section{One-loop calculation}

\begin{figure}[h!]
   \centering
   \includegraphics[scale=.5]{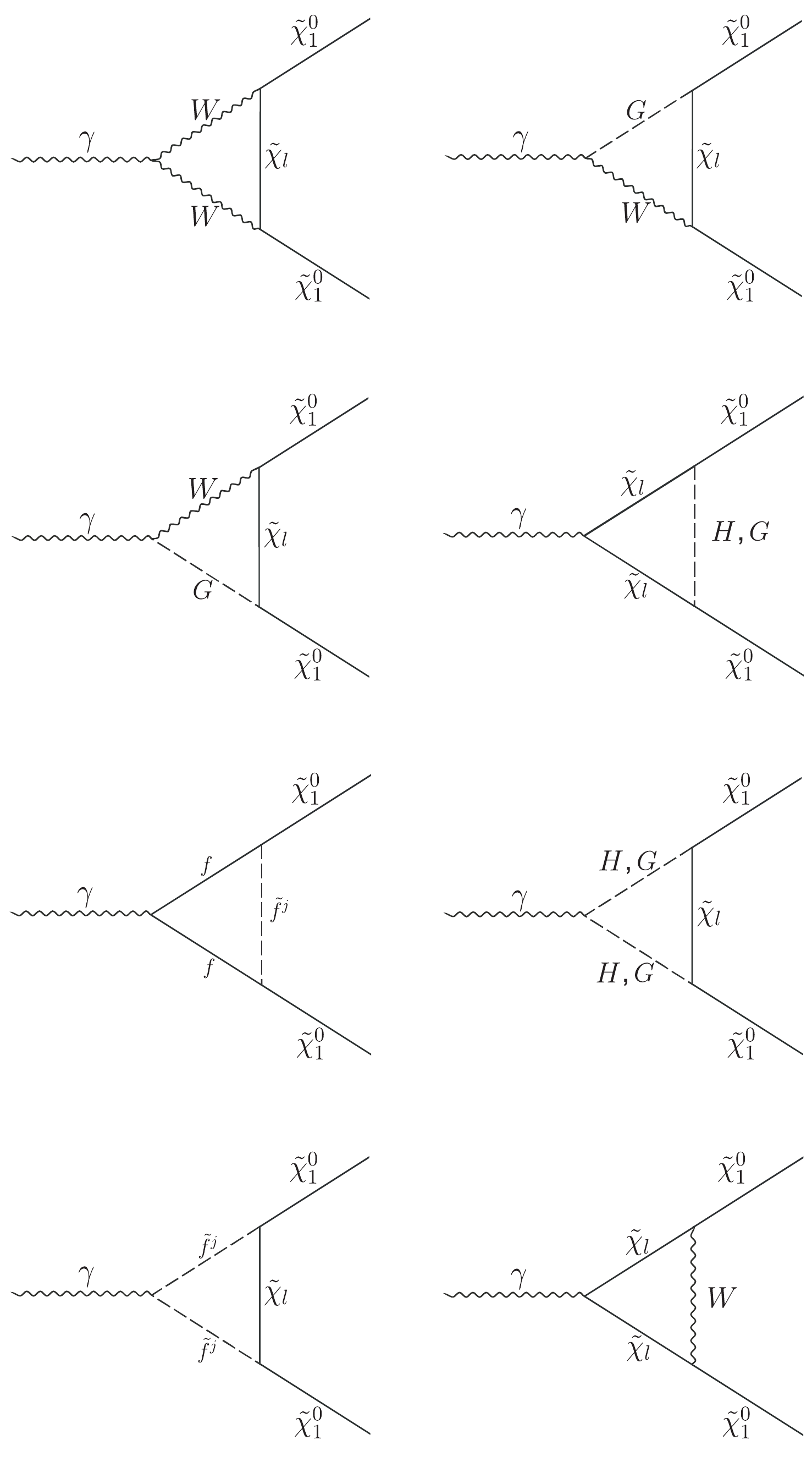}
   \caption{One-loop vertex corrections to the process
$\gamma\longrightarrow \chi_{1}^{0} \chi_{1}^{0}$.}
\label{one-loop-vertex}
\end{figure}

\begin{figure}[h!]
   \centering
   \includegraphics[scale=.2]{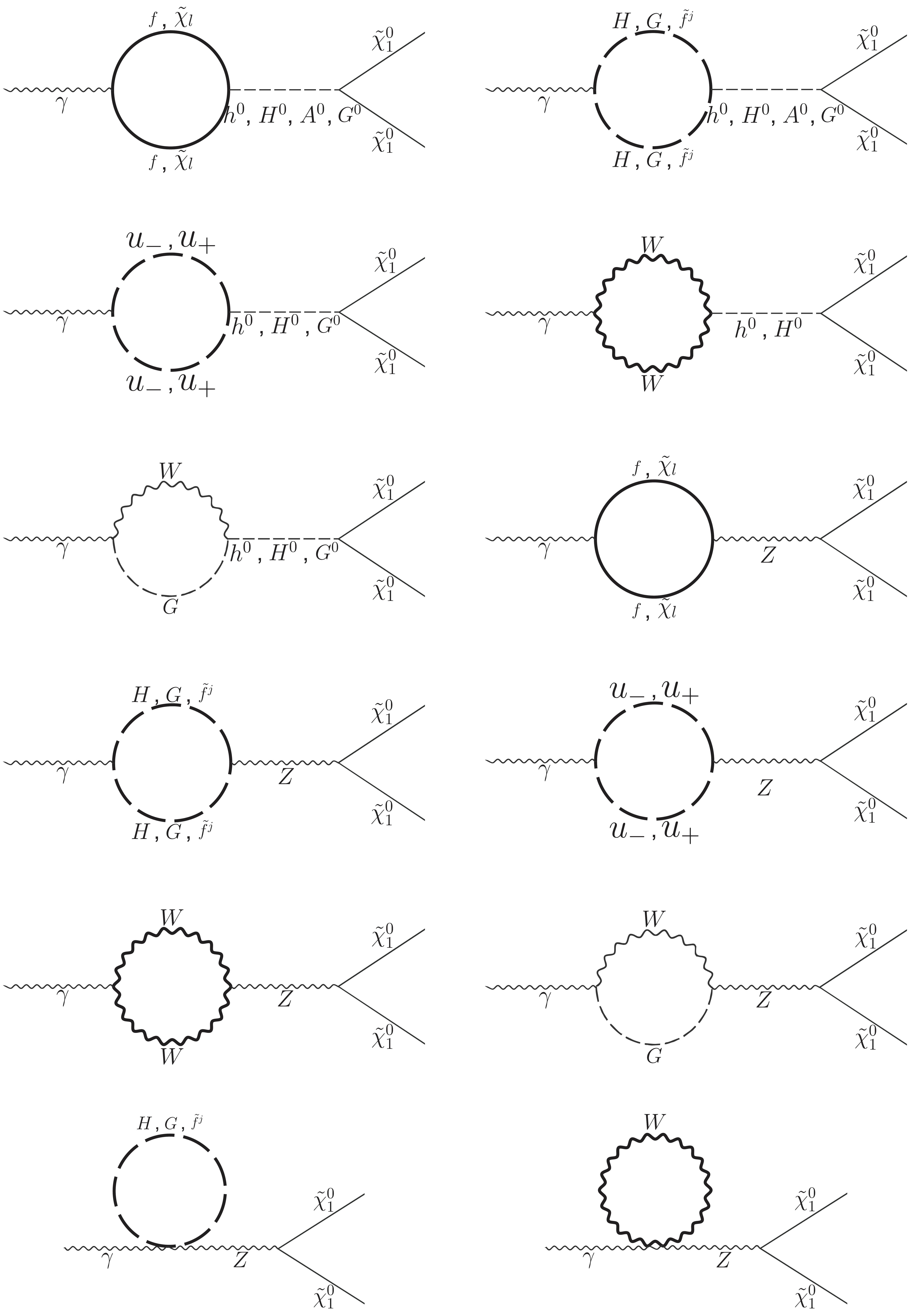}
   \caption{One-loop corrections to the self-energy for the process
     $\gamma\longrightarrow \chi_{1}^{0} \chi_{1}^{0}$.}
\label{one-loop-self}
\end{figure}

A neutralino is a Majorana particle, and it is necessarily electrically
neutral. This fact does not allow for a tree level electromagnetic
coupling. Therefore the electromagnetic properties of the Majorana
particle --the anapole-- arise only via loop contributions. The
anapole moment of the neutralino may be defined in the one-loop
approximation in the MSSM by the Feynman diagrams shown in
Figs.~\ref{one-loop-vertex} and \ref{one-loop-self}, where $f$
represents the charged fermions of the SM.  Taking each fermionic
family separately we obtain 96 Feynman diagrams in total,
corresponding to self-energies and vertex corrections.

We use \textit{FeynCalc} \cite{Mertig:1990an} to calculate the amplitude of these diagrams.
Since we are only interested in the terms that contribute to the
anapole form factor, we isolate the ones that have the Lorentz
structure $\gamma_{\mu}\gamma_5$. One of the first results we
obtain is that the self-energies $\gamma H^{0}$, $\gamma h^{0}$,
$\gamma A^{0}$ and $\gamma G^{0}$ do not contribute to the
calculation at all.  If we call  $\Xi_{i}$ the coefficient that multiplies 
$\gamma_{\mu}\gamma_5$ for the  \textit{ith} diagram, then we have
that
\begin{equation}
\sum_{i} \Xi_{i} = f_A (q^2 )q^2.
\end{equation}

To obtain the anapole moment $a = f_A(0)$ 
we use the l'Hopital rule and get
\begin{equation}
  a = f_A(0) = \lim _{q^{2}\rightarrow 0} \frac{\sum_{i} \Xi_{i}}{q^2} =
 \frac{\partial \sum_{i} \Xi_{i}}{\partial q^2} \mid _{q^{2}\rightarrow0}.
\label{anapole-hopital}
\end{equation}
Two- and three-point Passarino-Veltman scalar functions arise in the
calculation of each diagram. The two-point PV scalar function is
defined as
\begin{equation}
B_0 (q^2; m_1^2 , m^2_2) \equiv \frac{(2 \pi \mu)^{4- D}}{i \pi^2}
\, \int \frac{d^D k}{[k^2 - m_1^2]
[(k + q)^2 - m_2^2]} \, ,
\end{equation}
and the three point PV scalar function is defined as
\begin{equation}
\begin{array}{c}
\displaystyle
C_0 (A, B, C; m_1^2, m_2^2, m_3^2) \equiv
\\[0.5cm]
\displaystyle
\frac{(2 \pi \mu)^{4- D}}{i \pi^2}
\int \frac{d^D k}{[k^2 - m_1^2]
[(k + p_1)^2 - m_2^2][(k + p_2)^2 - m_3^2]} \, .
\end{array}
\end{equation}
where $A = p_1^2, B = (p_1 - p_2)^2$ and $C = p_2^2$. The
self-energies contain two point Passarino-Veltman scalar functions of
the type $B_{0} \left( q^{2}, x^2, x^2 \right)$ and $B_{0} \left( 0,
  x^2, x^2 \right)$.  Likewise, the contributions to the vertex
corrections have two and three point scalar functions of the type
$B_{0} \left( q^{2}, x^{2}, x^{2}\right)$, $B_{0} \left(
  m^{2}_{\tilde{\chi} _{1}^{0}}, y^{2}, x^{2}\right)$ and $C_{0}
\left( q^2, m^{2}_{\tilde{\chi} _{1}^{0}}, m^{2}_{\tilde{\chi}
    _{1}^{0}}, x^{2}, x^{2}, y^{2} \right)$.  In both cases $x$ and
$y$ represent the masses of the particles in the loop.

When evaluating (\ref{eq:electro-neutralino}), derivatives of the
Passarino-Veltman functions appear. To evaluate the $B_{0}$s, as well
as their derivatives, we use \textit{LoopTools}\cite{Hahn:1998yk}. To
evaluate the $C_{0}$s and their derivatives we expand them in a power
series around $q^2 =0$.   In this way it is possible to
  find an analytic approximation which coincides with the full
  expression in the limit $q^2 =0$, simplifying enormously the calculation (see appendix).

\begin{figure}[t]
   \centering
   \includegraphics[scale=.4]{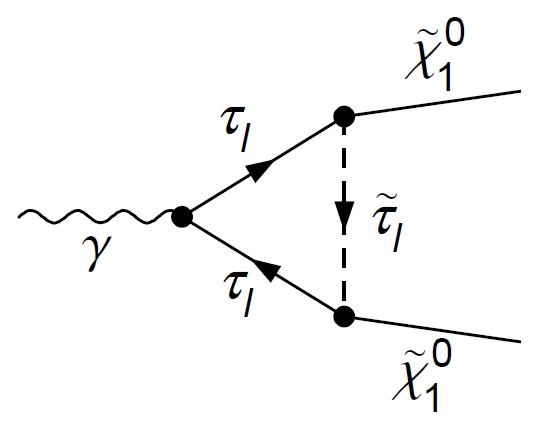} \hspace{1cm}
   \includegraphics[scale=.4]{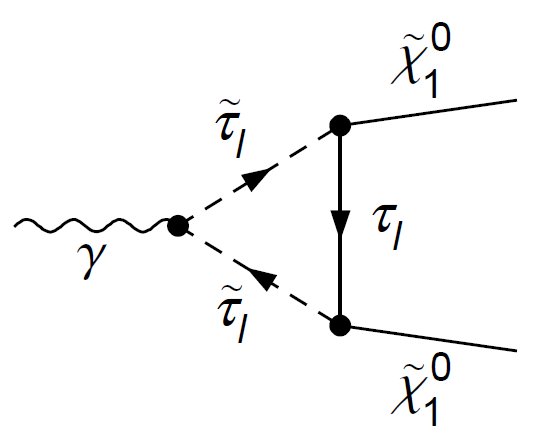}
   \caption{Two of the four dominant Feynman diagrams for the calculation of the anapole moment of the neutralino. The other two are equal but with internal arrows directed counterclockwise.}
\label{stau2}
\end{figure}

  In all regions of parameter space (except for $m_{1/2} \gg m_0$,
  which is ruled out by cosmological constrains since the LSP is
  charged) it was observed that the four triangle diagrams involving
  $\tilde{\tau}$ in the loop are almost completely
  dominant. (Fig.~\ref{stau2} shows two of these diagrams. The other
  two are equal, but with the flow arrows going counterclockwise due
  to the Majorana nature of the neutralino.) The approximate analytical expressions for the
  contributions of these diagrams are
\begin{eqnarray}
 \Xi_{1} \approx \frac{-k}{q^2-4m^{2}_{\tilde{\chi} _{1}^{0}}} \left\{ (q^2-2m^2_{\tilde{\tau}}+2m^{2}_{\tilde{\chi} _{1}^{0}})B_0 (q^2, m^2_{\tau}, m^2_{\tau}) +2(3m^{2}_{\tilde{\chi} _{1}^{0}}-m^2_{\tilde{\tau}})B_0(m^{2}_{\tilde{\chi} _{1}^{0}}, m^2_{\tau},m^2_{\tilde{\tau}}) \right. \\ \nonumber
\left. +2\left[(m^{2}_{\tilde{\chi} _{1}^{0}}-m^2_{\tilde{\tau}})^2-q^2m^{2}_{\tilde{\chi} _{1}^{0}}\right] C_0(q^2, m^{2}_{\tilde{\chi} _{1}^{0}}, m^{2}_{\tilde{\chi} _{1}^{0}},  m^2_{\tau}, m^2_{\tau}, m^2_{\tilde{\tau}}) - (q^2-4m^{2}_{\tilde{\chi} _{1}^{0}}) \right\}
\end{eqnarray}
and
\begin{eqnarray}
\Xi_{2} \approx \frac{k}{q^2-4m^{2}_{\tilde{\chi} _{1}^{0}}} \left\{ (q^2-2m^2_{\tilde{\tau}}-2m^{2}_{\tilde{\chi} _{1}^{0}})B_0 (q^2, m^2_{\tilde{\tau}}, m^2_{\tilde{\tau}})+2(m^{2}_{\tilde{\chi} _{1}^{0}}-m^2_{\tilde{\tau}})B_0(m^{2}_{\tilde{\chi} _{1}^{0}}, m^2_{\tau},m^2_{\tilde{\tau}}) \right. \\ \nonumber
\left. -2(m^{2}_{\tilde{\chi} _{1}^{0}}-m^2_{\tilde{\tau}})^2 C_0(q^2, m^{2}_{\tilde{\chi} _{1}^{0}}, m^{2}_{\tilde{\chi} _{1}^{0}}, m^2_{\tilde{\tau}}, m^2_{\tilde{\tau}}, m^2_{\tau}) + (q^2-4m^{2}_{\tilde{\chi} _{1}^{0}}) \right\},
\end{eqnarray}
where $k$ is given by
\begin{eqnarray}
k = \frac{e^3 \left[ Z_{1,1}(Z_{2,1} +Z_{1,2})\cos \theta_W \sin \theta_W +Z_{2,1}Z_{1,2} \cos^2 \theta_W-3Z^2_{1,1}\sin^2 \theta_W \right](\cos^2\theta_{\tilde{\tau}}-\sin^2\theta_{\tilde{\tau}})}{128 \pi^2 \cos^2 \theta_W \sin^2 \theta_W (q^2-4m^{2}_{\tilde{\chi} _{1}^{0}})}.
\end{eqnarray}
These expressions depend on the masses of the particles involved; $Z_{i,j}$, the elements of the neutralino mixing matrix; $\theta_{\tilde{\tau}}$, the $\tilde{\tau}$ mixing angle; as well as the electroweak angle $\theta_W$.

We evaluate the anapole moment within the cMSSM using \textit{Suspect}
\cite{Djouadi:2002ze}, by fixing the value of $A_0$, $\tan \beta$ and
${\rm sign}\mu$, and scanning over the other two parameters, $m_0$ and
$m_{1/2}$. 
We then vary $A_0$, $\tan \beta$, and ${\rm sign}\mu$, and repeat the
procedure. These values we then input into our own code to calculate
the anapole moment.  This code includes all diagrams contributing to
the anapole moment, evaluated using the approximation given in the appendix for the
Passarino-Veltman $C_0$ functions in the limit
$q^2 \to 0$.  The expressions for each diagram are not shown
explicitly in this paper.

We have not considered in our analysis the region where
  $m_{\tilde{\chi^0_1}}$ and $m_{\tilde{\tau}}$ are degenerate or quasi-degenerate, since
  this corresponds to a pole in the anapole moment function.  Other
  methods than the one we used here should be employed to analyse this
  region.
 Despite the fact that ${\rm sign}\mu > 0$ may solve the
problem of the discrepancy between the measured value of $g-2$ of the
muon and the one predicted by the SM, this does not mean negative
sign$\mu$ is ruled out since this problem might be solved through
other mechanisms, therefore sign$\mu<0$ should be also taken into
consideration.  

\begin{figure}
  \centering
  \includegraphics[width=8cm]{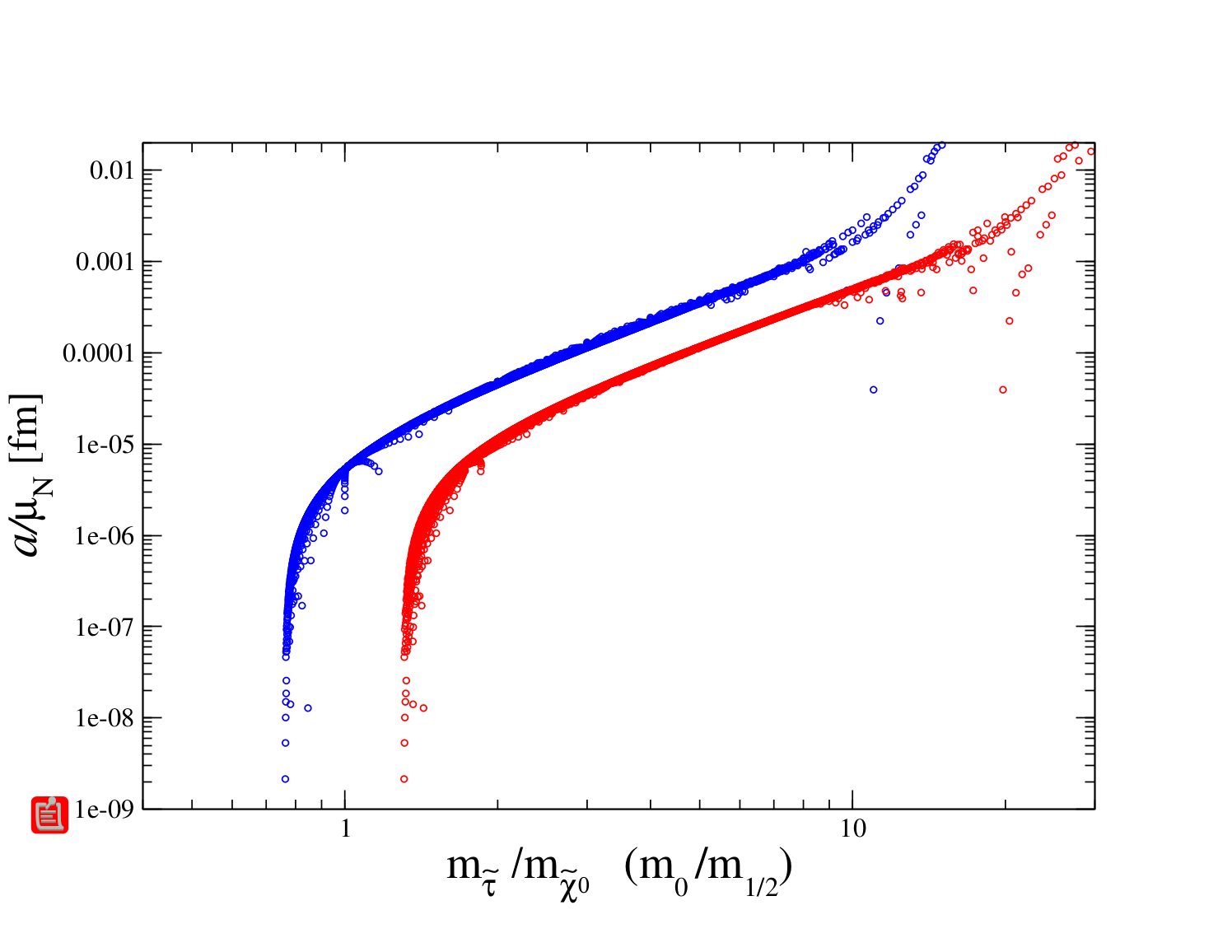}
  \caption{Dependence of the anapole moment (per nuclear magneton)
    with the  $m_{\tilde{\tau}}/m_{\tilde{\chi}^0_1}$ (red) and $m_0/m_{1/2}$ (blue) ratios, for
    $\tan\beta = 50$. We show only the values where the LSP is the lightest neutralino, and not the stau. The dependence for $\tan\beta=10$ is similar.}
\label{fig:anapole_vs_ratios}
\end{figure}

\begin{figure}[t]
   \centering
   \includegraphics[scale=.21]{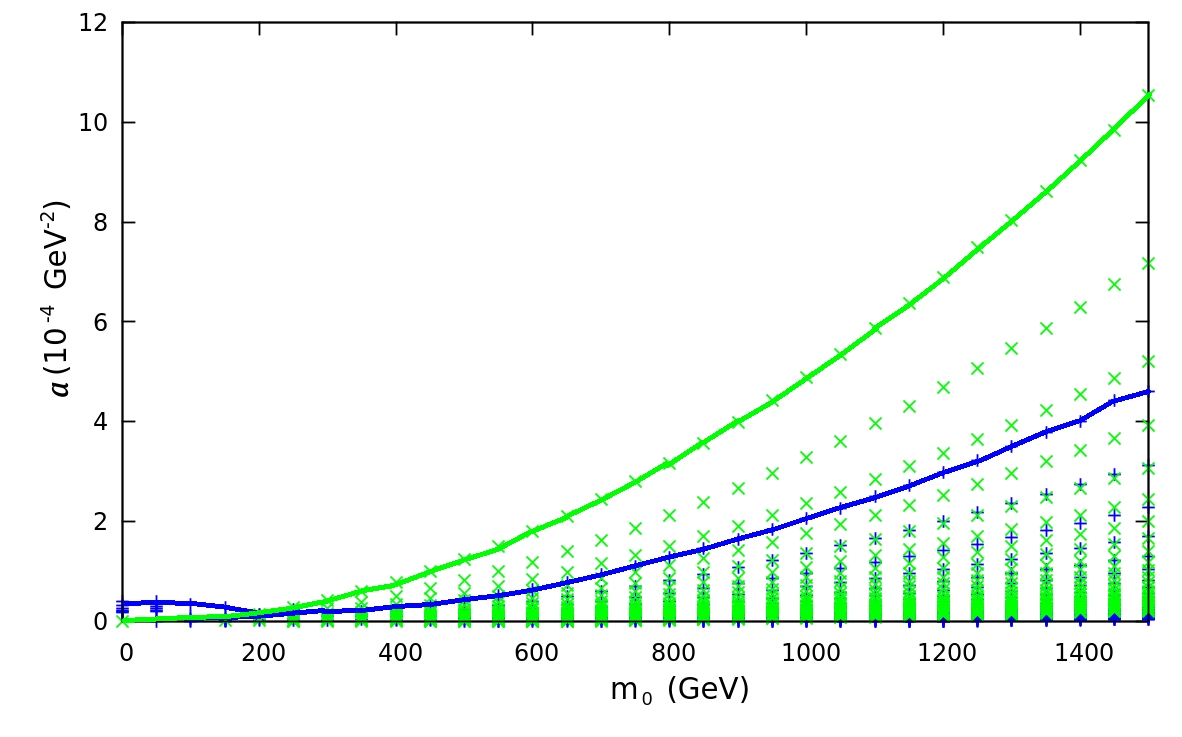} 
   \includegraphics[scale=.21]{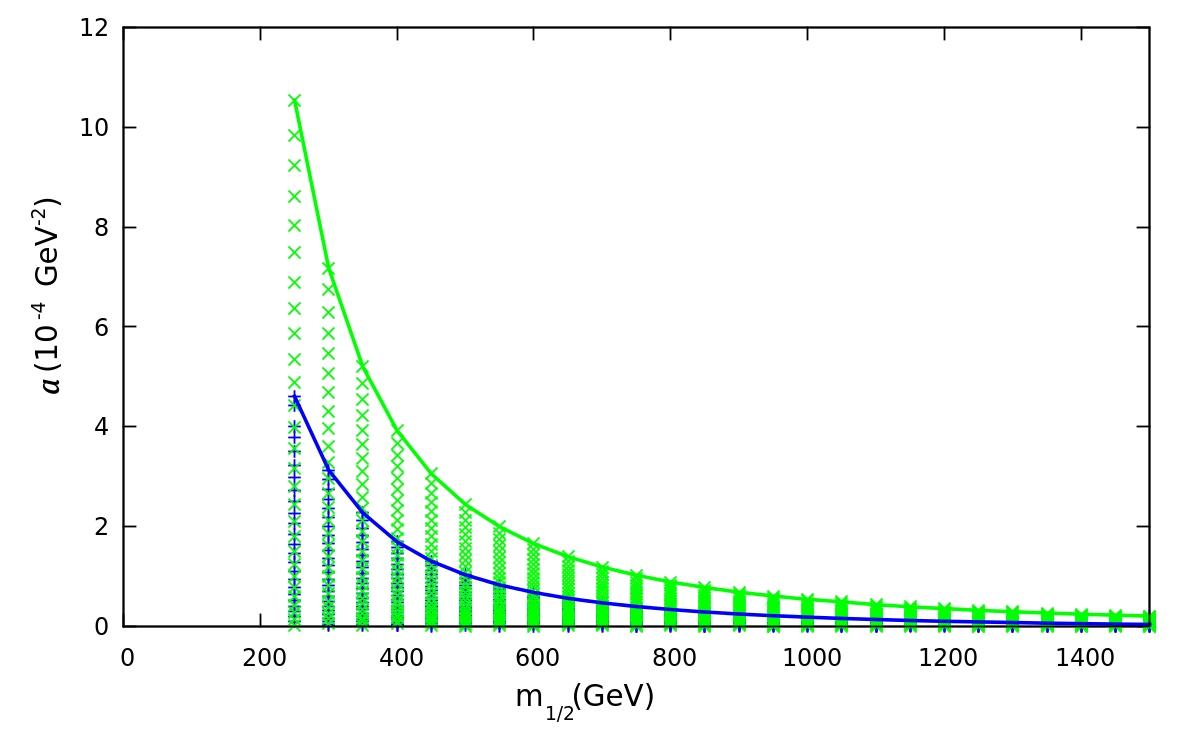}
   \caption{Anapole moment as a function of $m_0$ and $m_{1/2}$ ($A_0
     = 0$ and ${\rm sign}\mu > 0$ for $\tan \beta= 50$ (blue) and
     $\tan \beta= 10$ (green). The left panel shows the $m_0-a$ plane
     projection while the right panel shows the $m_{1/2}-a$ one. The
     solid line gives the maximum value of $a$ for that particular
     $\tan \beta$.}
\label{tan10-50}
\end{figure}

Although the full expression for the anapole moment is a complicated
function of the various parameters, it depends mainly on the relative
values of $m_{\tilde{\chi^0_1}}$ and $m_{\tilde{\tau}}$, which in turn
depend on the ratio of $m_0/m_{1/2}$, as can be seen in
Fig.~\ref{fig:anapole_vs_ratios}.
No dependence on $A_0$ or ${\rm sign}\mu$ was found,  however,
the anapole moment does depend slightly on $\tan \beta$ as can be seen in
Fig.~\ref{tan10-50}. 

%
%

Cosmological and experimental constrains have highly reduced the
allowed regions of parameter space of the cMSSM. The most recent
of these constrains is the one from the CMS and ATLAS collaborations
on the mass of the lightest Higgs boson $m_h = 125.8 \pm 0.6$ GeV 
\cite{Aad:2012tfa,ATLAS:2013mma,Chatrchyan:2012ufa,Chatrchyan:2013lba}. This
constraint is derived from a combination of $5.1$ fb$^{-1} \sqrt{s} =
7$ TeV data and $12.2$ fb$^{-1} \sqrt{s} = 8$ TeV data.  In our
calculation we take this constraint as $m_h = 126 \pm 3$, where the
uncertainty comes from a combination of the experimental and
theoretical determinations of the Higgs mass.
 
There is also a new measurement of \beq BR( \tilde{B}_s \rightarrow
\mu ^+ \mu ^-) = (3.2 \pm 1.5) \times 10^{-9} \eeq from the LHCb
collaboration, derived from $1 {\rm fb}^{-1}$ of data at $\sqrt{s} =
7$ TeV collision energy and $1.1 {\rm fb}^{-1}$ of data at $\sqrt{s} =
8$ TeV collision energy \cite{CMSLHCb}. The excluded region due to
this constraint in the cMSSM has already been determined in
\cite{Arbey:2012ax}. We also impose the constraint coming from the
branching ratio of $b\to s\gamma$, whose value is given by the Heavy
Flavour Averaging Group (HFAG) as~\cite{bsgexp}
 \beq
 \brat(b \to s \gamma ) = (3.55 \pm 0.24
{}^{+0.09}_{-0.10} \pm 0.03) \times 10^{-4} .
\label{bsgaexp}
\eeq
This result can be determined directly form Suspect.  Moreover, if we
assume that neutralinos make up all of the dark matter in the
universe, the WMAP 7-year dark matter relic abundance value
$\Omega_{\chi}h^2 = 0.1109 \pm 0.0056$ \cite{Larson:2010gs} puts even
more strict constraints. We calculated this value using micrOMEGAs
\cite{Belanger:2006is}. In general, the ``surviving'' regions have a
  very small value for the anapole moment of the lightest neutralino
  ($10^{-6}$-$10^{-7}$ GeV$^{-2}$), which is consistent with the
  results of Ho and Scherrer \cite{Ho:2012bg}.

\begin{figure}
  \includegraphics[width=12cm]{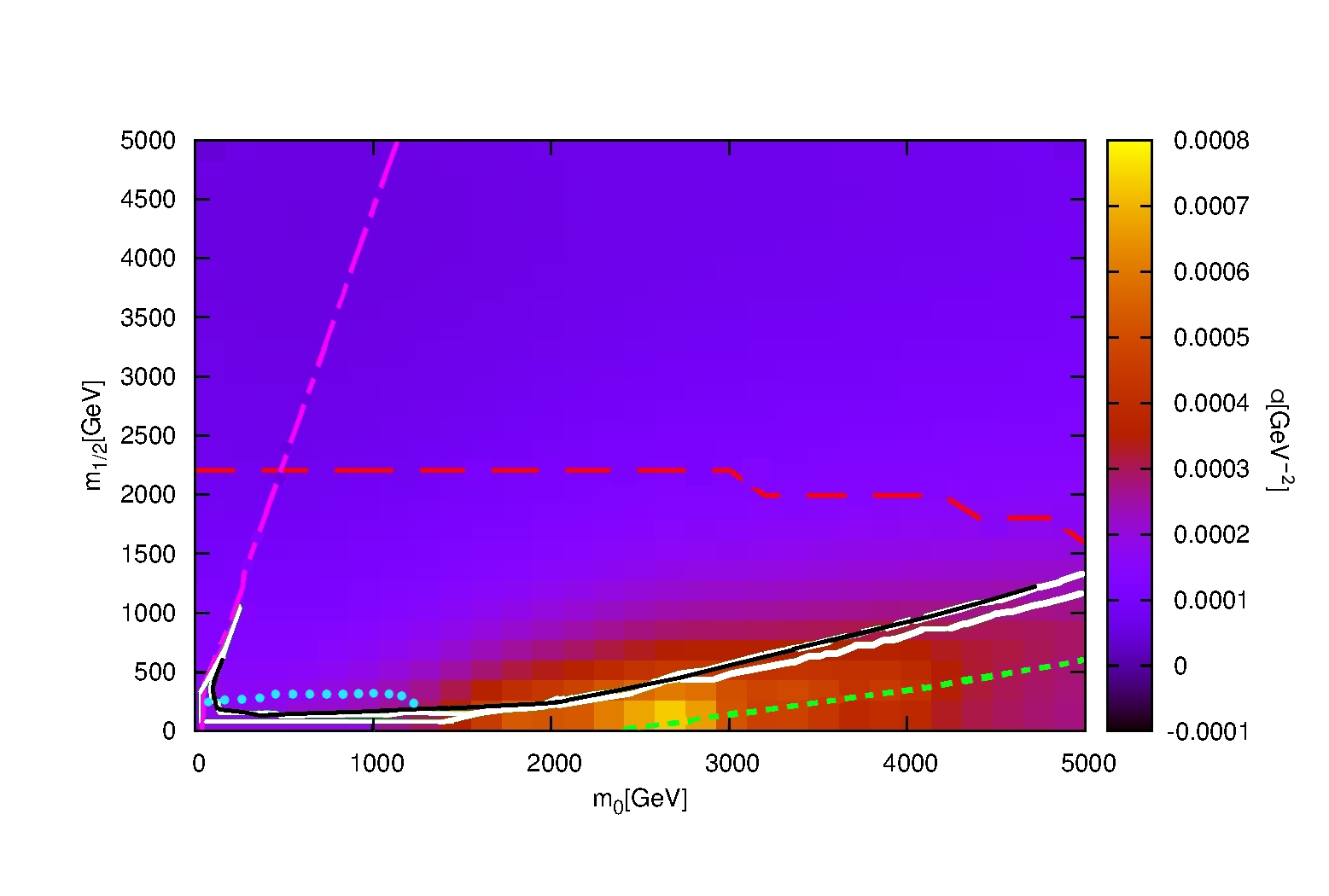}\\
   \includegraphics[width=12cm]{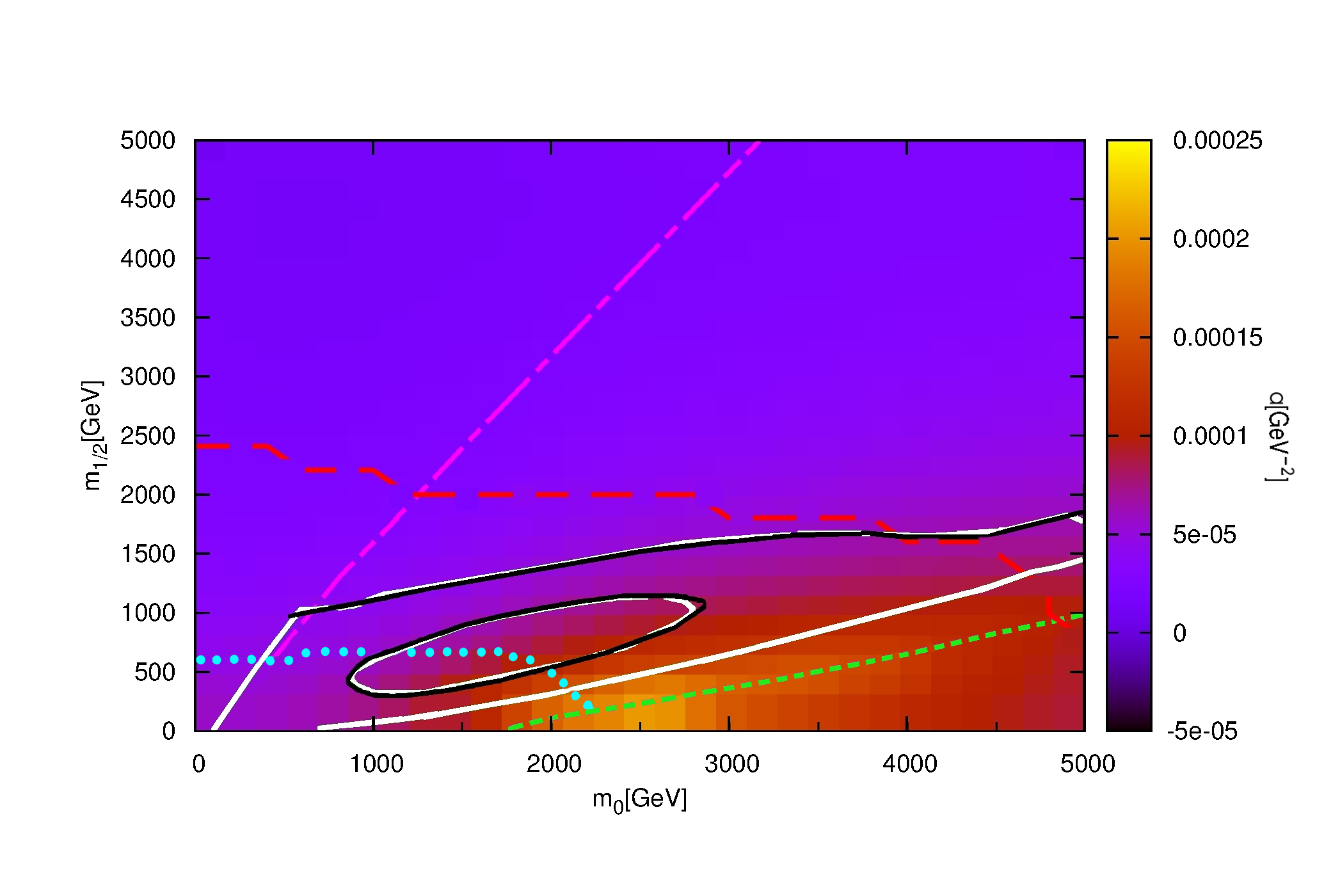}
   \caption{Value of the anapole moment in ($m_0, m_{1/2}$) planes in
     the cMSSM for $\tan \beta = 10$ (upper plot) and $\tan \beta =
     50$ (lower plot), assuming $A_0 = 0$ and $\mu > 0$, see text.
     The region that complies with all the phenomenological constraints 
      would be to the extreme right of the
     plots, between the white lines.}
\label{planes1}
\end{figure}

In the upper plot of Fig. \ref{planes1} we show the anapole moment
values for different regions of parameter space in ($m_0, m_{1/2}$)
planes in the cMSSM for $\tan \beta = 10$. On top and closely around
the pink dot-dashed line the stau and neutralino masses are
degenerate, and we do not calculate the anapole moment in this
region. The plotting program extrapolates between the values on both
sides of the line, but there is actually a gap in the data there.

  The
different phenomenological constraints are shown as follows: the region above
the red dashed line is where the Higgs mass is $m_h = 126 \pm 3$ GeV,
to the left of the pink dot-dashed line the LSP is charged, under
the dotted blue line is the region excluded by the value of
$b\to s\gamma$, whereas the region under the dotted green line
is excluded because it does not comply with the requirement of radiative
electroweak symmetry breaking.  The region where the relic LSP density
falls within the range allowed by WMAP  is marked with a white
line, while a more loose constraint, $\Omega_{\chi}h^2 < 0.12$,
assuming the LSP is not the only component of CDM, is delimited by a
white line.  The lower plot shows the same regions, but with
$\tan\beta = 50$.

In the two graphs we can see that the anapole moment is $\lesssim
{\cal O}(10^{-4})$GeV$^{-2}$ for every region of the space of
parameters. That is the case for the bulk (low $m_0$ and low
$m_{1/2}$), already excluded by other constraints. This is also true
for the coannihilation region (low $m_0$ and higher $m_{1/2}$), where
the masses of the $\tilde{\chi} _{1}^{0}$ and the $\tilde{\tau}$ are
almost degenerate. There seems to be a mechanism, which is a function
of $m^{2}_{\tilde{\chi} _{1}^{0}} - m^2_{\tilde{\tau}}$, suppressing
the contributions from this region. As can be seen from the approximate analytical formulae, there is a dependence on the mass difference between the stau and the neutralino both in the numerator and the denominator. It has to be remembered though that the anapole moment depends on the derivatives of these expressions, which obscures the mechanism at hand. The anapole moment gets relatively
large in the focus point region, which corresponds to high $m_0$ and
low $m_{1/2}$, close to the border of viable EWSB (green dotted
line). The same behavior is seen for different $\tan\beta$, although
the regions might differ in position and size.

\begin{figure}
  \includegraphics[width=10cm]{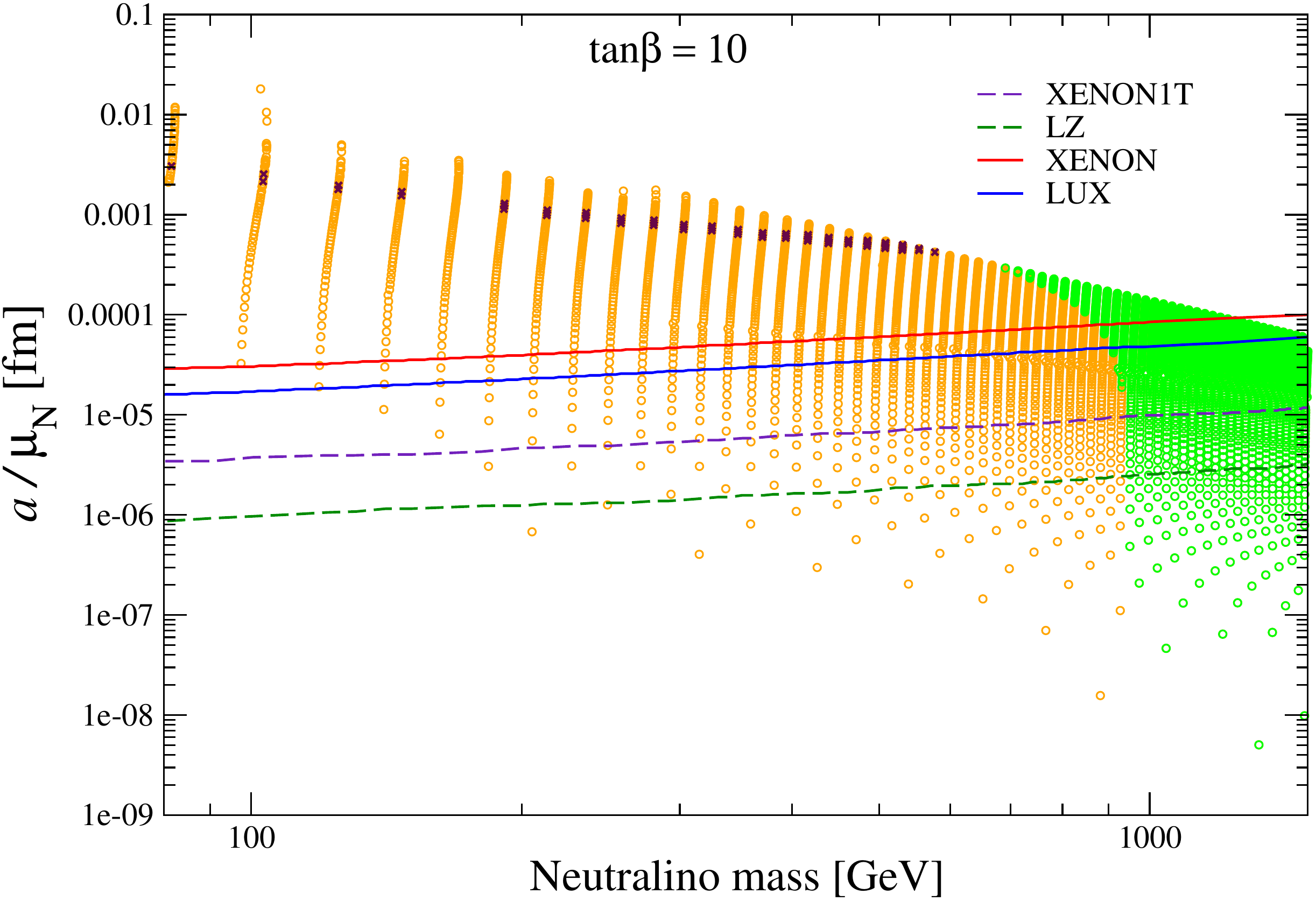}
  \includegraphics[width=10cm]{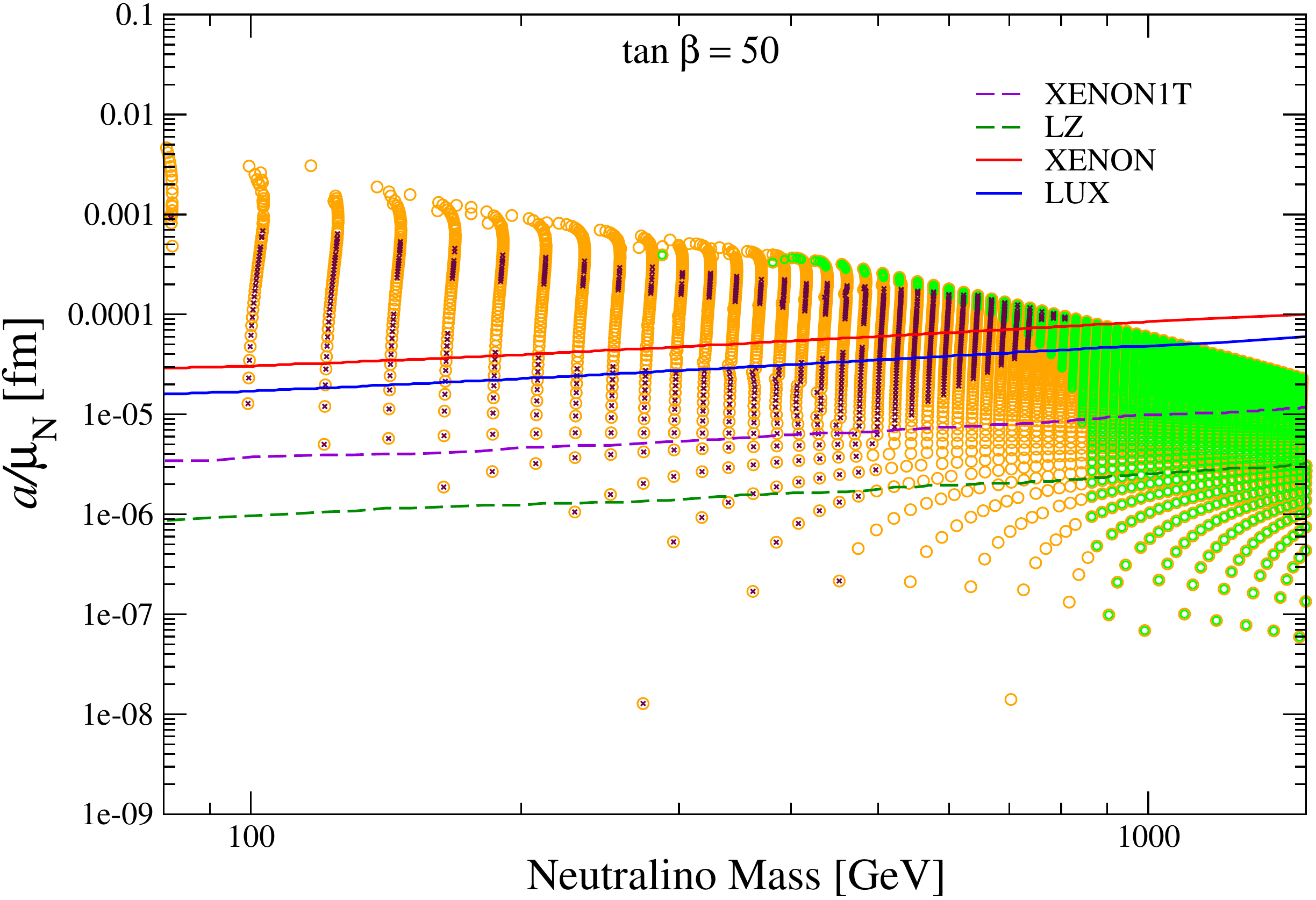}
  \caption{The plots show the dependence on the anapole moment (per
    nuclear magneton) on the neutralino mass for $\tan\beta = 10$ and
    $50$.  The regions above the solid lines correspond to the
    excluded regions by XENON and LUX, and the ones above the dashed
    lines correspond to the projected reach of XENON1T and LZ. (See text for an explanation of the different regions.)}
\label{fig:anapole_vs_neu-4exp}
\end{figure}

In Fig.~\ref{fig:anapole_vs_neu-4exp} we show the anapole
  moment dependence on the neutralino mass, and the exclusion curves
  of XENON and LUX \cite{Kopp:2014tsa}, plus the projected reach of
  XENON1T and LZ \cite{Ibarra:2016dlb}.  The regions above the lines
  of LUX and XENON are the values already excluded by the
  non-observation of dark matter.  The orange points are all the ones
  where the lightest neutralino is the LSP, the brown ones are the
  ones that comply with the loose relic density constraint, and the
  green ones comply with the range of values we took for the Higgs
  mass.  As can be seen from the graphs, the region in parameter space
  that complies with the Higgs mass constraint is at the border of
  the reach of LUX and XENON, but within the reach of LZ and
  XENON1T. For $\tan\beta=10$ there is practically no region where
  there is an overlap between the loose relic abundance and Higgs mass
  constraints, so this region is basically excluded by all three
  constraints (relic abundance, Higgs mass, and lack of observation of
  dark matter).  For $\tan\beta=50$ there is a small overlap region
  between the Higgs mass and relic density constraints.  This region
  lies at the border of the exclusion region of LUX and XENON, but
  within the projected reach of LZ and XENON1T. If in the future it is
  possible to measure the anapole moment of a DM candidate, this would give us extra information on the allowed parameter space of the cMSSM.

The cMSSM may be too constrained to be realistic, however, using it
  as a test model, we can see that the anapole moment is indeed
  different for different regions of parameter space, and is within the reach of the future
  experiments. Thus, anapole analysis can be used as another criteria
  to study the parameter space. Although the
  anapole moment is insensitive to $A_0$ and sign$\mu$, other
  observables, like the Higgs mass and some decays, are not
  \cite{Strege:2012bt,Fowlie:2012im,Roszkowski:2014wqa,Buchmueller:2012hv,Buchmueller:2013psa,Bagnaschi:2015eha}.
  This will give different exclusion regions for different values of
  the cMSSM parameters.


\section{Conclusions}

We calculated the anapole moment of the lightest neutralino in the
framework of the cMSSM. Even though this is perhaps an unrealistically
constrained model, it is one of the most studied SUSY models, and therefore it is important to pass it through all possible tests. 

We found that the anapole
moment of the neutralino is sensitive to $m_0$, $m_{1/2}$ and $\tan
\beta$, but non-dependent on $A_0$ and sign$\mu$. The parameter space we scanned gives rise to an anapole moment
consistent with the upper limit  obtained by
Pospelov and ter Veldhuis \cite{Pospelov:2000bq} for WIMPs interacting
with heavy nuclei using data from the CDMS and DAMA experiments.
The experimental constraints (Higgs mass, CDM relic
  density) favour scenarios with large $\tan\beta$. For $\tan\beta=50$
  we found that the anapole moment of the lightest neutralino of the cMSSM has a value ${\cal O} (10^{-5} -
  10^{-4})$ fm $\mu_N$, which lies at the border of sensitivity of the current
  experimental searches and within the reach of the future experiments
  like XENON1T and LZ.

The same kind of calculation can be performed in the context of other more complex models. This could be extremely
valuable for discriminating not only between different interesting regions of the cMSSM, but also among different models. 


Thus, the anapole analysis could be useful to study the allowed
parameter space of the cMSSM and other SUSY models.


\section*{Acknowledgements}
We acknowledge very useful discussions with S. Heinemeyer, E. Ley Koo, and 
A. Mondrag\'on, and invaluable technical help from T. Bernal, C. Espinoza, and L. Nellen. We also thank the referee for very useful suggestions and comments that improved the paper.\\
This work was partially supported by UNAM grants PAPIIT IN113412 and
IN111115, and Conacyt grant 132059.

\section*{Appendix: Scalar Three-point Function}

In this appendix, we analyse the Passarino-Veltman scalar three-point
function $C_0(q^2,x^2,x^2,z^2,z^2,y^2)$ \cite{Cabral:2000, Cabral:2006} which appears in the TDM
calculation. Here $q^2$ denotes the photon transfered 4-momentum, $x$
is the neutralino mass, and $y$ and $z$ are the masses of the
particles running in the loop.

\begin{figure}[h]
   \centering
   \includegraphics[scale=.3]{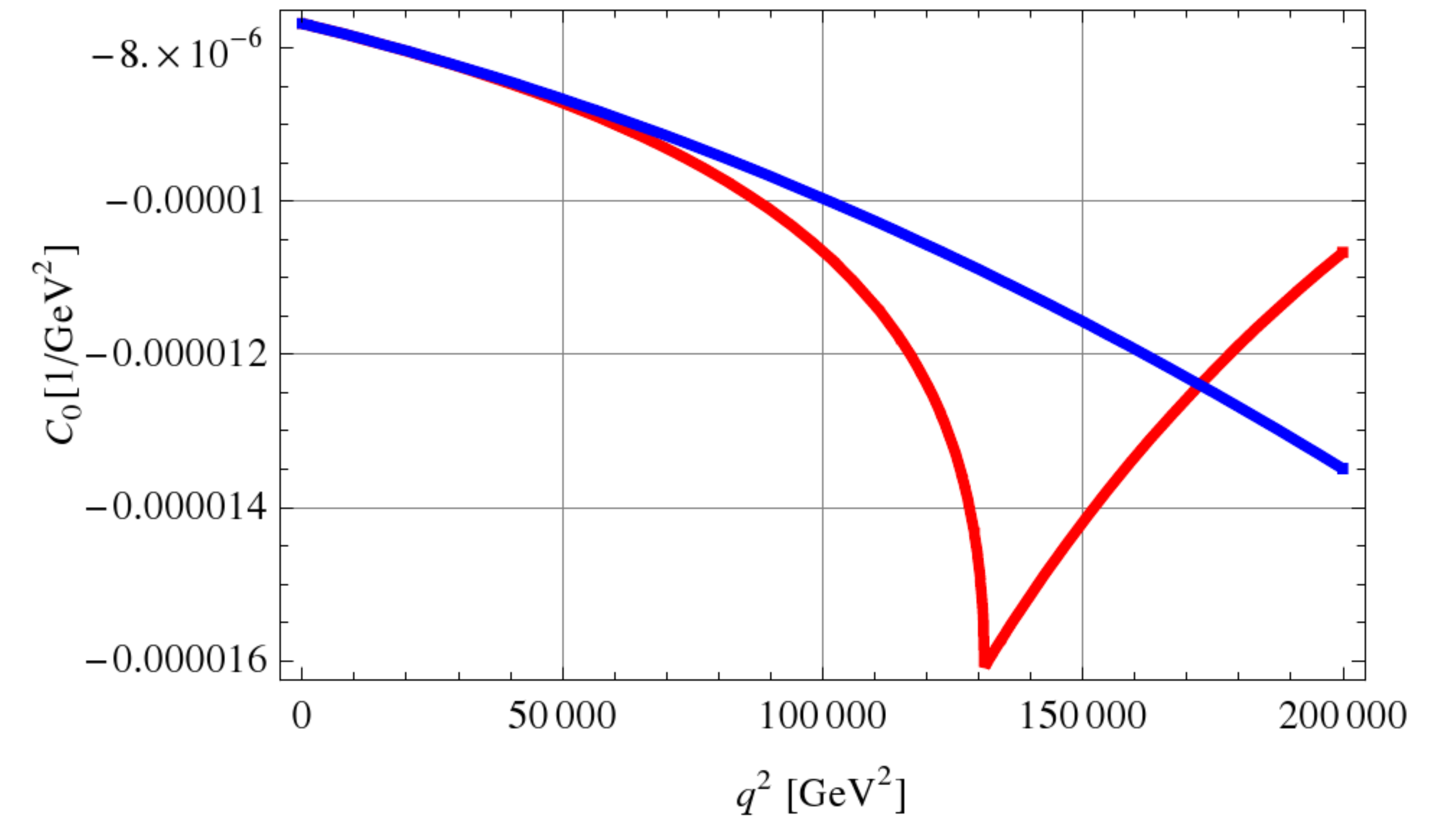}
   \caption{Comparison between numerical (red line) and approximate
     (blue line) scalar three-point function $C_0(q^2,x^2,x^2,z^2,z^2,y^2)$, with $x=97.7$ GeV, $y=415.4$ GeV and $z=80.43$ GeV. The analytical approximation (blue line) is only valid for $q^2\rightarrow 0$.}
\label{C0}
\end{figure}

The corresponding plot for this $C_0$ function can be seen in Fig. \ref{C0}.
The red line shows the numerical solution, the blue line
line represents the approximate solution, i.e., the Taylor expansion
around $q^2 = 0$, which can be written as follows:

\begin{equation}
C_0\left( q^{2},x^{2},x^{2},z^{2},z^{2},y^{2}\right) = \alpha_{0} +
\alpha_{1} q^{2} + {\cal O}(q^4).
\end{equation}

\noindent 
The coefficients $\alpha_i$ are functions of the masses:

\begin{equation}
\alpha_0 = \frac{\log\left( \frac{y^2}{z^2}\right)}{2 x^2} + a\log\omega, 
\end{equation}

\begin{equation}
\alpha_1 =
\frac{x^4-y^2x^2-2z^2x^2+z^4-y^2z^2}{6x^2z^2(-x+y-z)(x+y-z)(-x+y+z)(x+y+z)} 
+ \frac{\log\left( \frac{y^2}{z^2}\right)}{12 x^4}+ b\log \omega,
\end{equation}

\noindent where

{\scriptsize
\begin{equation}
\omega = \frac{\left(
  ix^2+iy^2-iz^2+\sqrt{-y^4+2(x^2+z^2)y^2-(z^2-x^2)}\right)
\left( ix^2-iy^2+iz^2+\sqrt{-y^4+2(x^2+z^2)y^2-(z^2-x^2)}\right) }
{\left( -ix^2+iy^2-iz^2+\sqrt{-y^4+2(x^2+z^2)y^2-(z^2-x^2)}\right)
\left( -ix^2-iy^2+iz^2+\sqrt{-y^4+2(x^2+z^2)y^2-(z^2-x^2)}\right)},
\end{equation}
}

\begin{equation}
a= \frac{i(x^2+y^2-z^2)}{2x^2\sqrt{-x^4+2y^2x^2+2z^2x^2-y^4-z^4+2y^2z^2}}
\end{equation}

\noindent and

{\scriptsize
\begin{equation}
b= \frac{i(x^2+y^2-z^2)(x^4-4y^2x^2-2z^2x^2+y^4+ź^4-2y^2z^2)}
{12x^4(-x+y-z)(x+y-z)(-x+y+z)(x+y+z) \sqrt{-x^4+2y^2x^2+2z^2x^2-y^4-z^4+2y^2z^2}}.
\end{equation}
}

\bibliography{biblio}
\bibliographystyle{h-physrev5}
\end{document}